\documentclass[fleqn,12pt,twoside]{article}
\usepackage{espcrc1}

\usepackage{graphicx}
\usepackage{epsfig}
\usepackage[figuresright]{rotating}

\newcommand{\AmS}{{\protect\the\textfont2
  A\kern-.1667em\lower.5ex\hbox{M}\kern-.125emS}}

\begin{document}

\def\E{\epsilon}

\title{Jet Quenching }

\author{R.~Baier\address{Fakult\"at f\"ur Physik,
 Universit\"at Bielefeld \\ 
        Postfach 10 01 31, D-33501 Bielefeld, Germany }
\thanks{Talk given at "Quark Matter 2002", Nantes, France, July 18-24, 2002}
 \thanks{Supported, in part, by DFG, contract FOR 339/2-1.} }

\maketitle

\begin{abstract}
A short summary of the physics underlying jet quenching is given.
\end{abstract}

\section{Introduction}

In August 1982 J.~D.~Bjorken published a preprint {\cite{ref0}}
on "Energy Loss of Energetic Partons in Quark-Gluon Plasma:
Possible Extinction of High $p_{\perp}$ Jets in Hadron-Hadron Collisions",
in which he discussed that high energy quarks  and gluons propagating through
quark-gluon plasma (QGP) suffer differential energy loss, and where he
further  pointed
out that as an interesting signature events may be observed in which the
hard collisions may occur such that one jet is escaping without
 absorption and the other is fully absorbed.

The arguments in this work have been based on elastic scattering of high 
momentum partons from quanta in the QGP, with a resulting 
("ionization") loss $-dE/dz \simeq \alpha_s^2 \sqrt{\epsilon}$,
with $\epsilon$ the energy density of the QGP. The loss turns out to be less 
than the string tension of $O(1~\rm{GeV/fm})$ \cite{thoma} .

However, as in QED, bremsstrahlung is another important source of
energy loss \cite{ref7}. Due to multiple (inelastic) scatterings and induced
gluon radiation high momentum jets and leading large $p_{\perp}$ hadrons
become depleted, quenched \cite{ref1} or may even become extinct.
 In \cite{ref8} 
it has been shown that a genuine pQCD process (Fig.~\ref{fig:rad})
is responsible for the dominant loss: after the gluon is radiated off
the energetic parton it suffers multiple scatterings in the medium.
Indeed, further studies 
 by \cite{ref9,ref10,ref11,ref12,ref15,ref16,ref17,ref18}
support this observation \cite{review}.

In the following I mainly concentrate on the influence of the medium-induced 
energy loss on the large $p_{\perp}$ leading hadron spectrum. 
More about the recent theoretical developments can be learned
from the contributions  to the 
"High Transverse Momentum" session at this conference \cite{talks}.

It is important to mention that
for the first time  large $p_{\perp}$ leading hadron data 
from $Au-Au$ collisions have been measured by 
the PHENIX and STAR Collaborations 
at RHIC \cite{ref3,ref3a,ref4}.

\section{pQCD medium-induced radiative energy loss}

\begin{figure}[htb]
\begin{minipage}[t]{75mm}
\vspace{1.5cm}
\psfig{figure=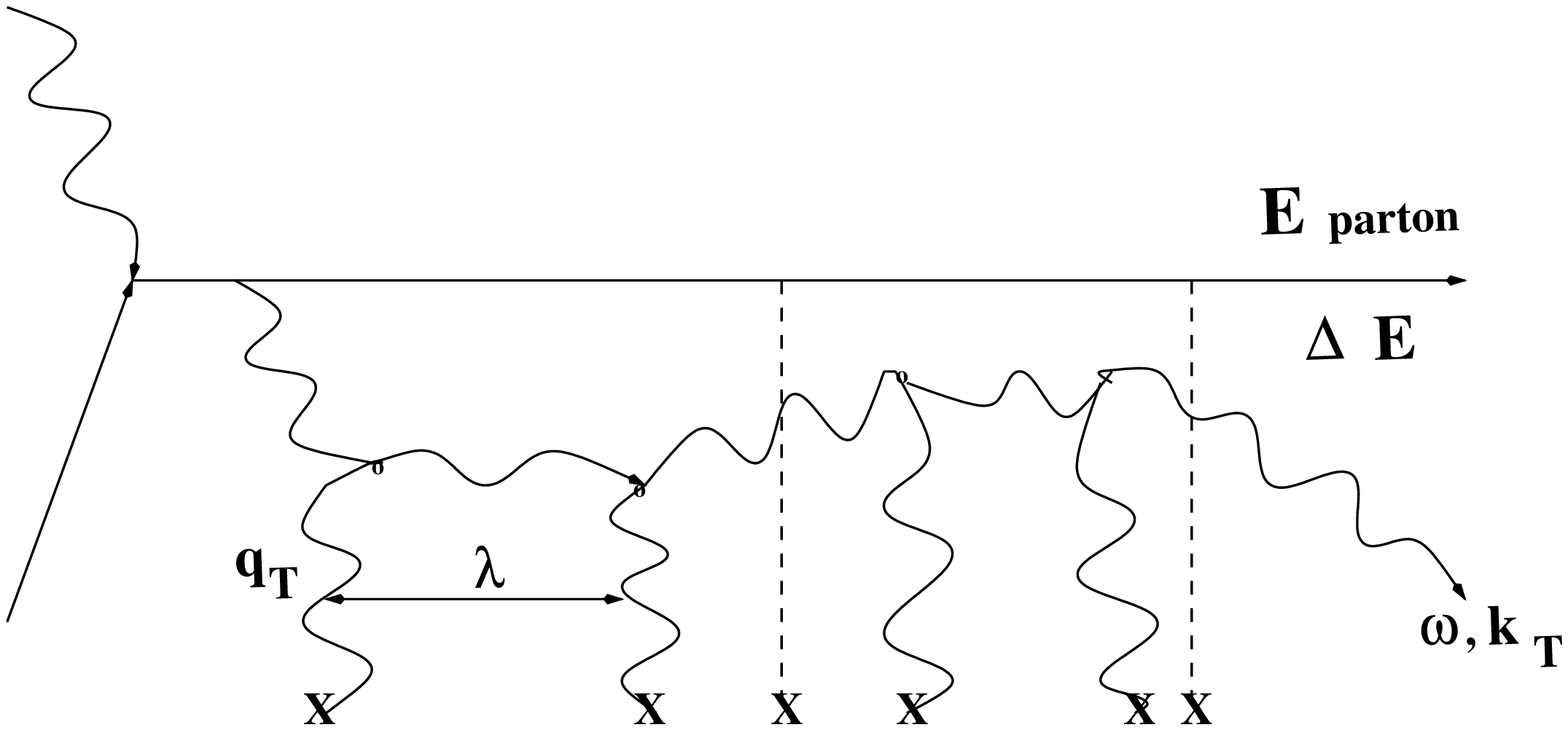,angle=-90,width=6.5cm}
\caption{Typical gluon radiation diagram }
\label{fig:rad}
\end{minipage}
\hspace{\fill}
\begin{minipage}[t]{75mm}
\psfig{bbllx=85,bblly=200,bburx=520,bbury=655,
file=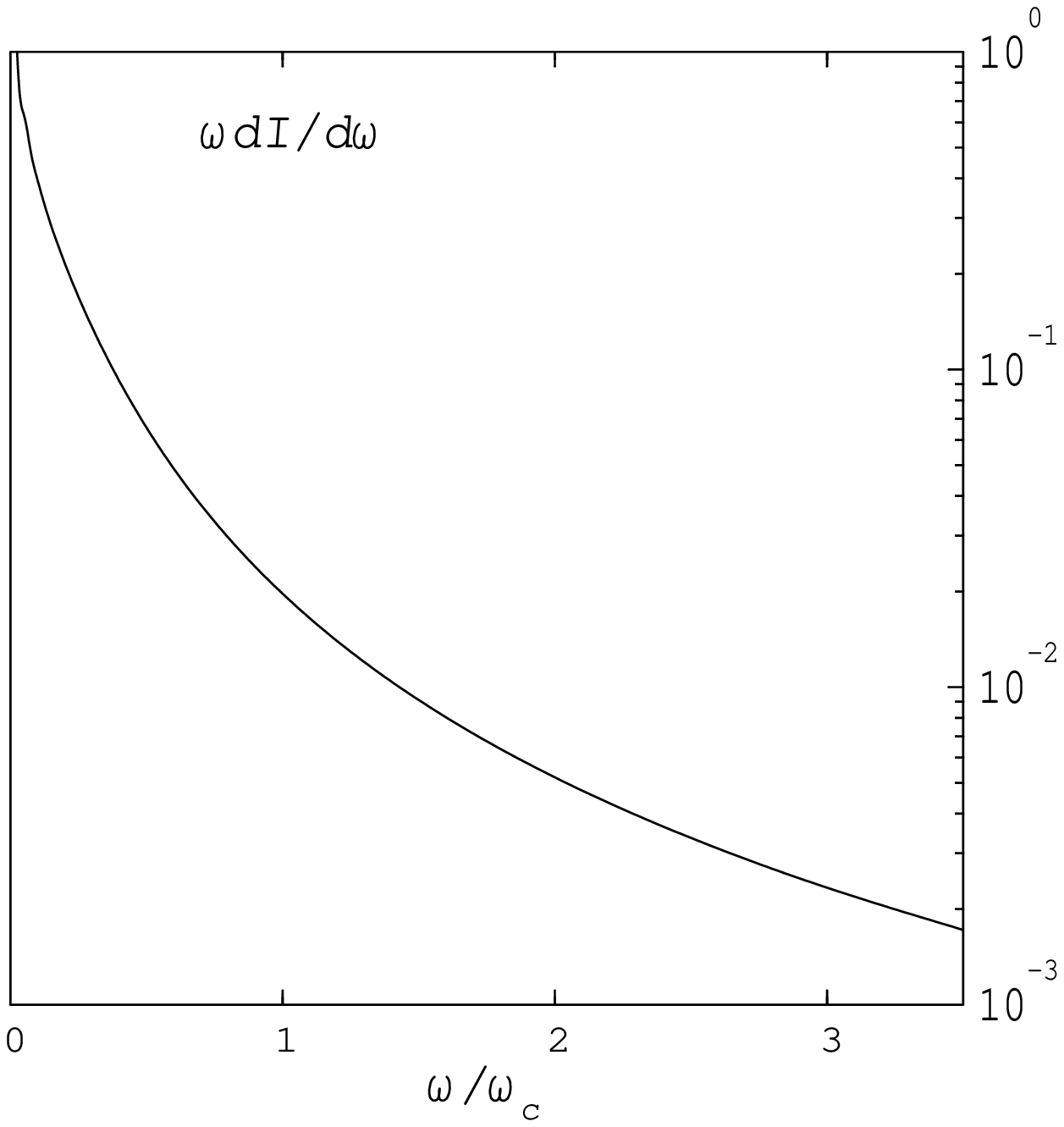,angle=-90,width=74mm}
\caption{Medium-induced soft gluon spectrum}
\label{fig:gluon}
\end{minipage}
\end{figure}

After its production in a hard collision the energetic parton
radiates a gluon which both traverse
a finite size $L$ medium.  Due to its non-abelian nature
and its interaction with the medium  this gluon
follows a zig-zag path (Fig.~\ref{fig:rad}),
with a  mean free path $\, \lambda~ > ~1/\mu~$, which is  the
 range of screened multiple gluon interactions.

It can be shown that the  average energy loss  of the parton 
(in the limit $E_{parton} \rightarrow \infty$)
due to gluon radiation with a spectrum $\frac{\omega dI}{d \omega}$
is determined by the characteristic gluon energy
$~\omega_c$ as follows,  
\begin{equation}
\Delta E = \int^{\omega_c}
\frac{\omega dI}{d \omega}~{d \omega} \simeq \alpha_s ~\omega_c \, \, , 
\label{loss}
\end{equation}
where
\begin{equation}
\omega_c = \frac{1}{2} \hat{q} L^2 \,  .
\label{omc}
\end{equation}
The medium dependence is controlled by the transport coefficient 
\begin{equation}
 \hat{q} \simeq \mu^2/\lambda \simeq \rho 
        \int~d^2 q_{\perp}~q^2_{\perp}~ d\sigma/d^2 q_{\perp} \, ,
\label{qhat}
\end{equation}
where $\rho $ is the density of the medium (a nucleus, or  partons)
 and 
 $\, \sigma $ the  cross section of the 
gluon-medium interaction.
In order to understand (\ref{loss}) 
the coherent pattern of the induced 
gluon radiation is important.

The following semiquantitative arguments  are crucial:

\vspace{0.3cm}

$\bullet \,$ number of coherent scattering centers, which act as 
a single source of gluon radiation:
$\, \, \,   N_{coh} \simeq t_{coh}/\lambda 
\simeq \sqrt{\omega/\mu^2 \lambda} >> 1$ 

$\bullet \,$ coherence/formation time: $\, \, t_{coh} \simeq
 \omega/k^2_{\perp}  \simeq \sqrt{\omega/\hat{q}}$

$\bullet \,$ random walk of emitted gluon (Fig.~\ref{fig:rad}): accumulated 
average transverse momentum
 $\, \, k^2_{\perp} \simeq N_{coh}~ \mu^2 ~ $

$\bullet \,$ medium-induced soft gluon spectrum 
in relation to the Gunion-Bertsch (GB) spectrum \cite{gunion}:
\begin{equation}
\frac{\omega dI}{ d \omega d z} \simeq  \frac{1}{N_{coh}}
\frac{\omega dI}{d \omega d z}\vert_{GB} \simeq  \frac{1}{N_{coh}}
\frac{\alpha_s}{\lambda} \simeq  
\alpha_s ~\sqrt{\hat{q}/\omega}  .
\label{spec}
\end{equation}

The medium-induced (BDMPS \cite{ref9,ref11}) gluon spectrum
(valid for finite size $\, L >> \lambda \, $ 
  and for soft gluon energies 
$\,\, \, \omega_{GB}=
\hat{q} \lambda^2 < \omega << E_{parton} \, \rightarrow \infty $)
with the characteristic behaviour:
$\, \,  \frac{\omega dI}{d\omega} \simeq
 \alpha_s ~\sqrt{\frac{\omega_c}{\omega}} \, $ , 
 $\, \omega < \omega_c~$,  
is shown in Fig.~\ref{fig:gluon}. It is suppressed by $1/N_{coh}$ for
$\omega > \omega_{GB}$ with
respect to the incoherent Gunion-Bertsch spectrum 
(cf. Eq.~\ref{spec}). For comparison with  QED the  LPM
suppressed photon spectrum behaves as $ \frac{\omega dI}{d\omega} \simeq
 \sqrt{\omega}$ \cite{LPM}.

\section{ Transport coefficient $\hat{q}$ }

The coefficient $\hat{q}$ (\ref{qhat}) can be calculated in terms
of the gluon structure function, i.e. 
for  nuclear matter, 
\begin{equation}
 \hat{q} =\frac{4 \pi^2 \alpha_s~N_c}{N_c^2 - 1}~\rho ~[x G(x,\hat{q}L )]
 \, ,
\end{equation}
where 
 $ xG(x, Q^2) $ is the gluon distribution for a nucleon and $\rho$
the nuclear density.

Gluon dynamics is responsible
for the following important relations:

{$\bullet  \, \,$ 
{relation to $q_{\perp}$ broadening}}

Due to multiple scatterings off nucleons there is 
transverse momentum broadening of the gluon, 
\begin{equation}
< q_{\perp}^2 > = \hat{q}~L \,  , 
\end{equation}
such that the differential energy loss is expressed by \cite{ref9} 
\begin{equation}
 - \frac{dE}{dz} \simeq  \alpha_s  < q_{\perp}^2 >  \, .
\end{equation}

{$\bullet  \, \,$ 
{relation to "saturation scale" $Q_s$ \cite{iancu}}}

The saturation momentum  of gluons for
central gluon-nucleus (radius $R_A$)  collisions 
at small $x$ is given by \cite{Mueller},
\begin{equation}
 Q_s^2 =  \frac{4 \pi^2 \alpha_s (Q_s)~N_c}{N_c^2 - 1}~\rho ~[x G(x,Q_s^2)]
 ~2 R_A \,  = 2 R_A~ \hat{q} \, .
\end{equation}

\vspace{0.3cm}

$\bullet  \, \ \,$ {medium dependence of $\hat{q}$}

\begin{figure}[htb]
\centerline{
\psfig{bbllx=25,bblly=210,bburx=500,bbury=630,
figure=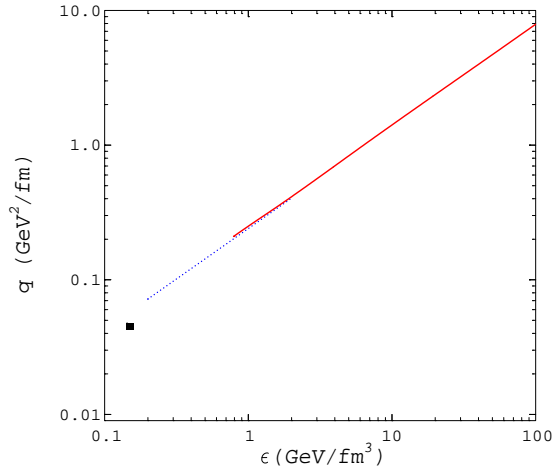,width=8cm} }
\caption{Transport coefficient as a function of energy density
for different media: cold, massless hot pion gas (dotted) 
and (ideal) QGP (solid curve) }
\label{fig:fig3}
\end{figure}

In Fig.~\ref{fig:fig3} the dependence of $\hat{q}$ as a function of the
 energy density of (equilibrated) media is shown; e.g. for QGP the number
density is translated into $\epsilon$ as
$\rho(T) \sim T^3 \sim \epsilon^{3/4}$. A "smooth" increase of $\hat{q}$
with increasing $\epsilon$ is  observed, such
that 

\begin{equation}
\hat{q} \vert_{\rm{QGP}} >> \hat{q} \vert_{\rm{nuclear~ matter}} \, .
\end{equation}

However, it is difficult to deduce from the behaviour indicated in
 Fig.~\ref{fig:fig3}, how the QCD phase transition near
$\epsilon \simeq 1~ {\rm {GeV/fm^3}}$ \cite{karsch} may be observed
by measuring jet quenching.

\vspace{0.3cm}

$\bullet  \, \ \,$ { expanding medium \cite{exp,vitev,wang,salg}}

In case of a dynamically expanding collision region the gluon
radiation spectrum, and the resulting energy loss
depend on an effective $\hat{q}\vert_{\rm{eff}}$,
equivalent to a static coefficient, which is obtained by

\begin{eqnarray}
    \hat{q}(L)\vert_{\rm{eff}} & = &
\frac{2}{L^2} \int_{\tau_0}^L ~ d\tau (\tau -\tau_0) ~\hat{q}(\tau)
\nonumber \\
& \simeq & \frac{2}{2 - \alpha} ~\hat{q}(L) \, \, 
\, {\rm for} \, \, \,  \tau_0 \rightarrow 0  ,
\end{eqnarray}
when
$ \hat{q}(\tau) =  \hat{q}(\tau_0)~ (\frac{\tau_0}{\tau})^{\alpha}$,
where $ \tau_0$ is the starting time of the expansion, which, however, may
not be very small.
 $\alpha = 1$  corresponds to { Bjorken's} longitudinal expansion
\cite{Bj}.

\vspace{0.3cm}
Which medium is actually probed by quenching?
According to the possible time behaviour of the hadronic system produced in
heavy ion collisions: Colour Glass Condensate in the initial state -
(non-equilibrated) quark gluon matter - QGP - mixed phase \cite{LM},
it is most likely that the hard probes propagate through different 
expanding, not necessarily thermalized, but dense gluonic media.

\section{How to "measure" $\Delta E(L)$ ? }

\begin{figure}[htb]
\begin{minipage}[t]{75mm}
\epsfig{file=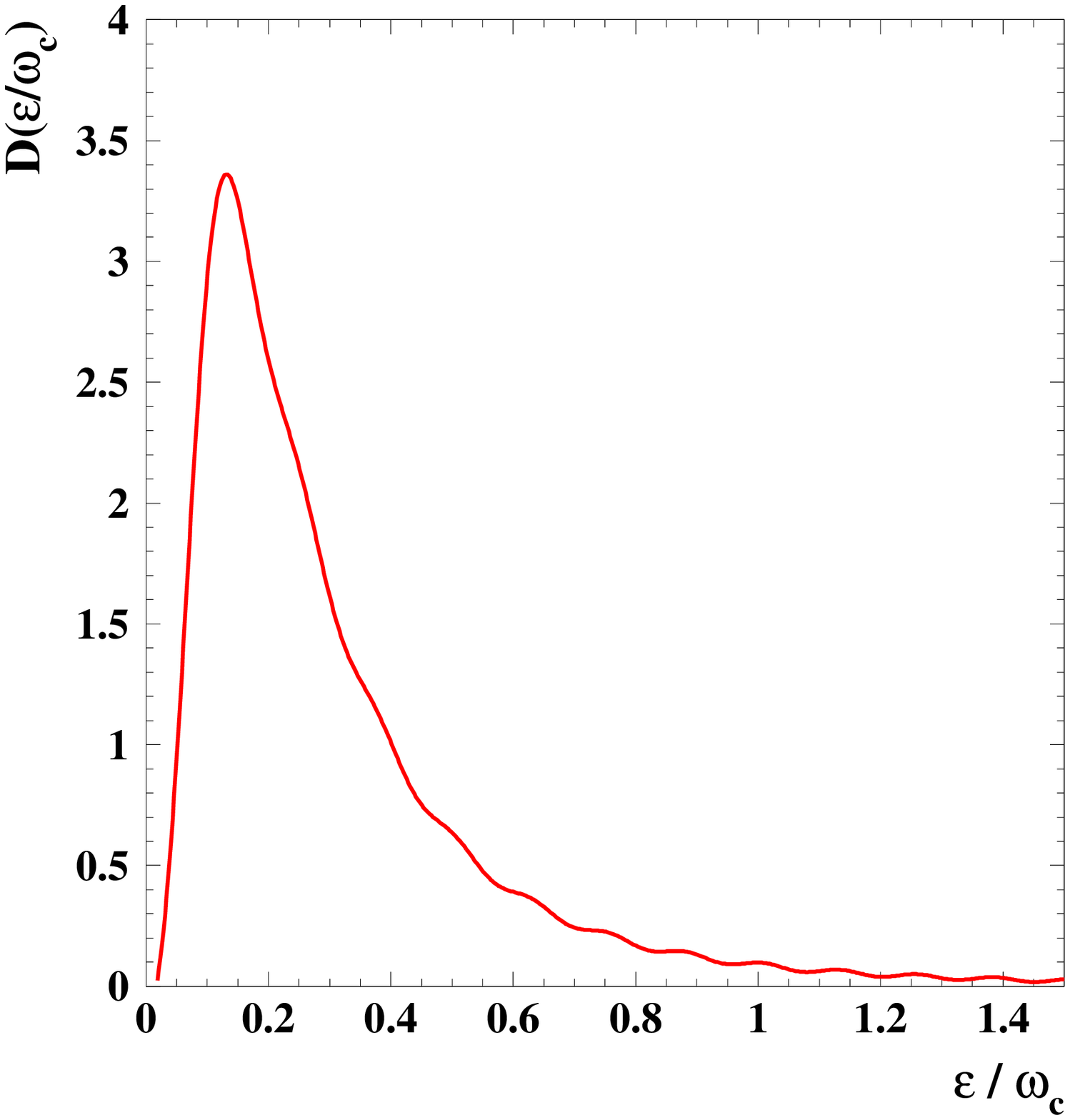, width=75mm}
\caption{Gluon radiation probability \cite{arleo}}
\label{fig:fig4}
\end{minipage}
\hspace{\fill}
\begin{minipage}[t]{70mm}
\epsfig{bbllx=100,bblly=190,bburx=490,bbury=640,
file=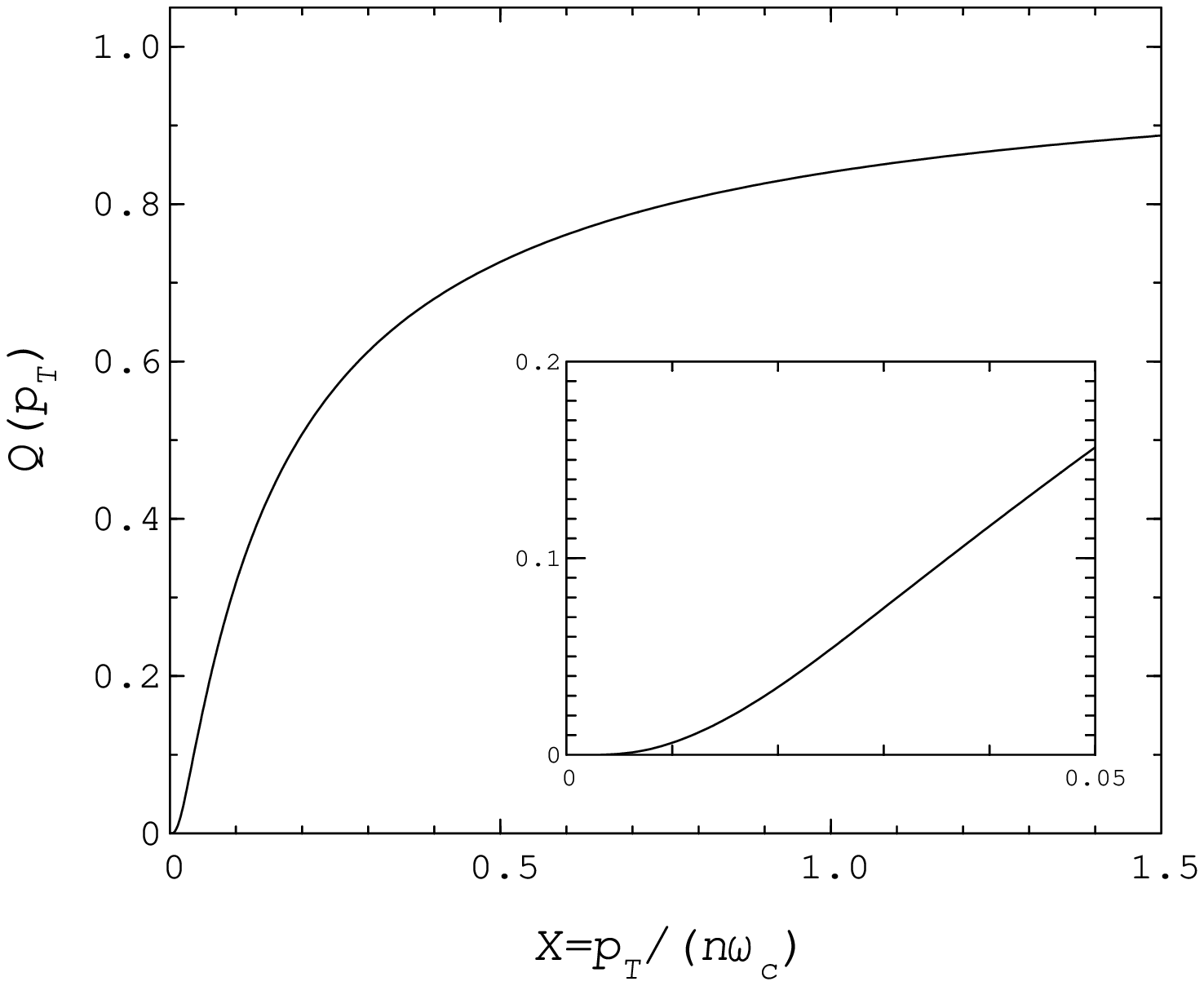, width=72mm}
\caption{The quenching factor $Q (p_{\perp})$  \cite{baier}}
\label{fig:fig5}
\end{minipage}
\end{figure}

\subsection{{ Quenching of leading hadron spectra in media  \cite{baier} }}

The yield of inclusive large $p_\perp$ hadrons in $A-A$ collisions
is essentially modified  due to the radiative medium induced energy loss,
leading to significant jet quenching, i.e. to a   shift of
the leading particle/pion spectrum

\begin{equation}
\frac{d\sigma^{{\rm medium}}(p_\perp)}{dp^2_\perp}  \simeq
\frac{d\sigma^{{\rm vacuum}}(p_\perp+S(p_\perp))}{dp^2_\perp} \, , 
\label{shift}
\end{equation} 
or, by introducing the medium dependent {\em quenching factor} 
$ Q(p_\perp) $,
\begin{equation}
\frac{d\sigma^{{\rm medium}}(p_\perp)}{dp^2_\perp} =
 \frac{d\sigma^{{\rm vacuum}} (p_\perp)}{dp^2_\perp}\cdot Q(p_\perp) \, , 
\label{quench}
\end{equation} 
which is related to the {\em shift} $S(p_\perp)$
by

\begin{equation}
 Q(p_\perp) = \exp\left\{-\frac{n}{p_\perp}\cdot
S(p_\perp)\right\} \, .
\end{equation}
The vacuum leading hadron distribution is a  steeply falling
 spectrum with
increasing  ${p_\perp}$,
 in terms of an effective power $n$,
\begin{equation}
\frac{d\sigma^{{\rm vacuum}}(p_\perp)}{dp_\perp^2} \propto \frac1{p_\perp^n}\,
, \qquad 
\,  \, \,  n=n(p_\perp)
   \>\equiv\> - \frac{d}{d\ln p_\perp} \> 
   \ln \frac{d\sigma^{{\rm vacuum}}(p_\perp)}{dp_\perp^2}   \, ,
\end{equation}
i.e. at RHIC:  $n \simeq 8 -10$.
This behaviour causes a strong bias leading to an additional suppression
 of real gluon emission, with the important result that the shift is
${p_\perp}$ dependent and 
that 
\begin{equation} 
S(p_\perp) < {\rm {average  \, \, loss}} \, \,    \Delta E \,  \,  !
\label{shiftrel}
\end{equation}

The {\it {additional}} energy loss due to medium induced gluon
radiation in the final state is characterized by the
probability $D(\E)$ that  radiated gluons carry away the
energy $\E$ by  
 independent emission of soft primary gluons,
\begin{equation}
D(\E) = \sum^\infty_{n=0} \, \frac{1}{n!} \,
\left[ \prod^n_{i=1} \, \int \, d\omega_i \, \frac{dI(\omega_i)}{d \omega} 
\right] \delta \left(\E - \sum_{i=1}^n  \omega_i\right)
\cdot \exp \left[ - \int d\omega \frac{dI}{d\omega} \right]  \, , 
\label{prob}
\end{equation}
where  $\frac{dI(\omega)}{d\omega}$ is the
 inclusive soft  LPM gluon spectrum given above (Fig.~\ref{fig:gluon}).
The last factor in (\ref{prob}) accounts for virtual corrections.
The distribution
$D(\E)$  peaks at small gluon energies
$ \E_{peak} <  \omega_c $ ,
as illustrated in Fig.~\ref{fig:fig4}.
To obtain the  inclusive hadron spectrum 
the vacuum  production cross section at energy $(p_\perp + \E)$
has to be convoluted with the distribution $D(\E)$,
\begin{equation} 
 \frac{d\sigma^{{\rm medium}}(p_\perp)}{dp^2_\perp} =
\int d\E \, D(\E) \, 
\frac{d\sigma^{{\rm vacuum}} (p_\perp + \E)}{dp^2_\perp} \, . 
\label{conv}
\end{equation}
The resulting quenching factor $Q(p_{\perp}) $ 
 scales in the  dimensionless  variable $X = p_{\perp}/(n \omega_c) \, . $
Fig.~\ref{fig:fig5} shows a 
significant suppression for small $X$ , i.e. 
for a { hot medium} $Q << 1 $. 
 Typical $X$ values are for $L = 5 ~ {\rm fm} \, ,  p_{\perp} = 10~ {\rm GeV},
n \simeq 10  :$ 
cold matter ($ \omega_c = 3~ {\rm GeV} $)  $X \simeq 0.3$ , 
hot matter ($ \omega_c = 25~ {\rm GeV}, ~ 
                   \epsilon = 2~ {\rm {GeV/fm^3}} $) $X \simeq 0.04 .$

\begin{figure}[htb]
\begin{minipage}[t]{80mm}
\epsfig{file=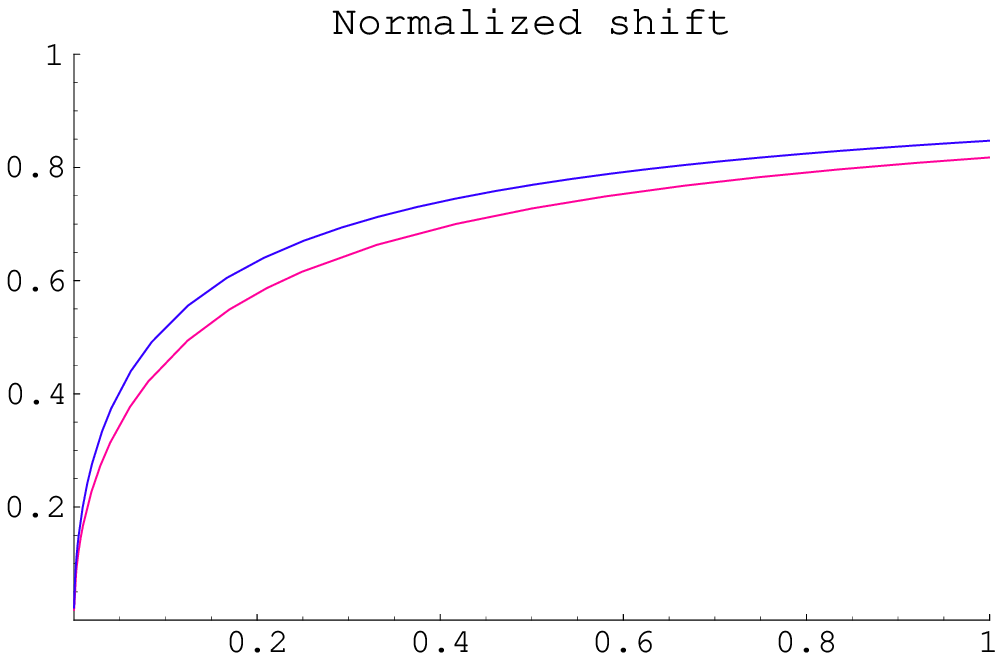,width=70mm}
\caption{Shift normalized to $\Delta E$
 as a function of $X$ for $n = 4$ (lower) and $n = 10$ (upper curve)
\cite{baier}}
\label{fig:fig6}
\end{minipage}
\hspace{\fill}
\begin{minipage}[t]{75mm}
\epsfig{file=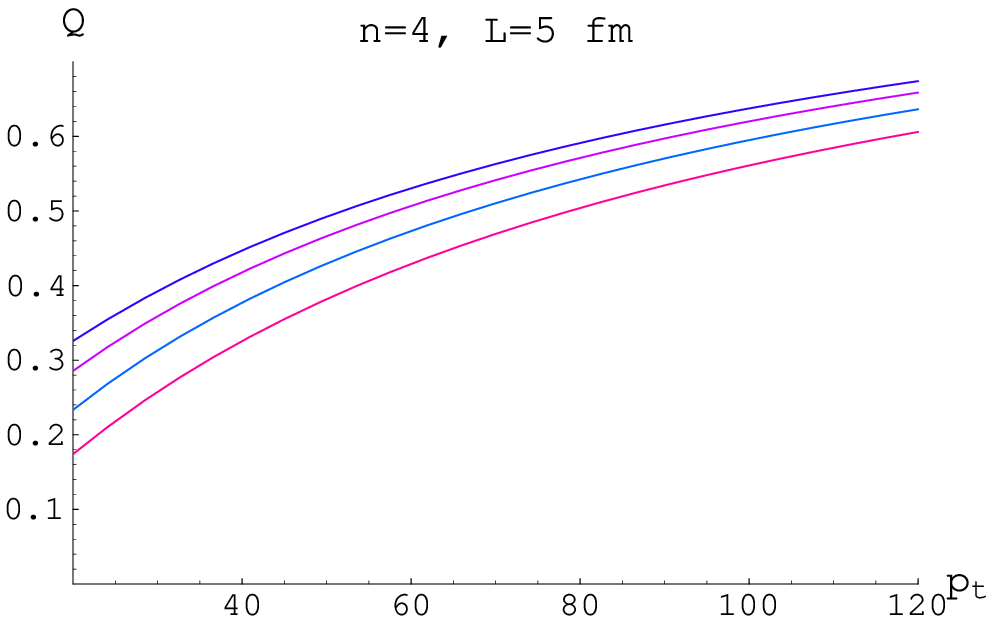,width=70mm}
\caption{"Infrared sensitivity" of Q
as a function of $p_{\perp}$ \cite{baier}}
\label{fig:fig7}
\end{minipage}
\end{figure}

In the region of interest for RHIC data the shift is plotted 
in Fig.~\ref{fig:fig6}, supporting the statement (\ref{shiftrel}),
and showing that effectively
$S(p_\perp) \simeq \alpha_s L \sqrt{{\hat q} p_\perp /n}$ \cite{baier}.

Although $Q(p_\perp)$ is formally an infrared-safe quantity,
the soft LPM spectrum $\propto 1/\sqrt{\omega}$ (Fig.~\ref{fig:gluon})
causes a serious instability for the range $ p_\perp < 10 ~{\rm GeV}$.
At the same time  in the large $p_\perp \ge 20 ~{\rm GeV}$ range
(where the characteristic gluon energies are much larger than $\omega_{GB}$)
the perturbative treatment for quenching is applicable,
meaning that the "infrared sensitivity" is not large.  
For an illustration 
this sensitivity  is shown in Fig.~\ref{fig:fig7}; the curves
from bottom to top 
correspond to  gluon energies $\omega$ cut by $
\omega \ge \omega_{cut}= 0,100,300,500$ MeV, respectively.

\subsection{\bf {Medium-modified fragmentation functions }}

Instead of (\ref{conv}) medium modifications may be studied
 more directly in the  fragmentation of partons $\rightarrow$ hadrons.

{$\bullet  \, \,$ 
{ model with gluon radiation probability $D(\epsilon)$ \cite{salg}}}

\begin{figure}[htb]
\begin{minipage}[t]{80mm}
\psfig{file=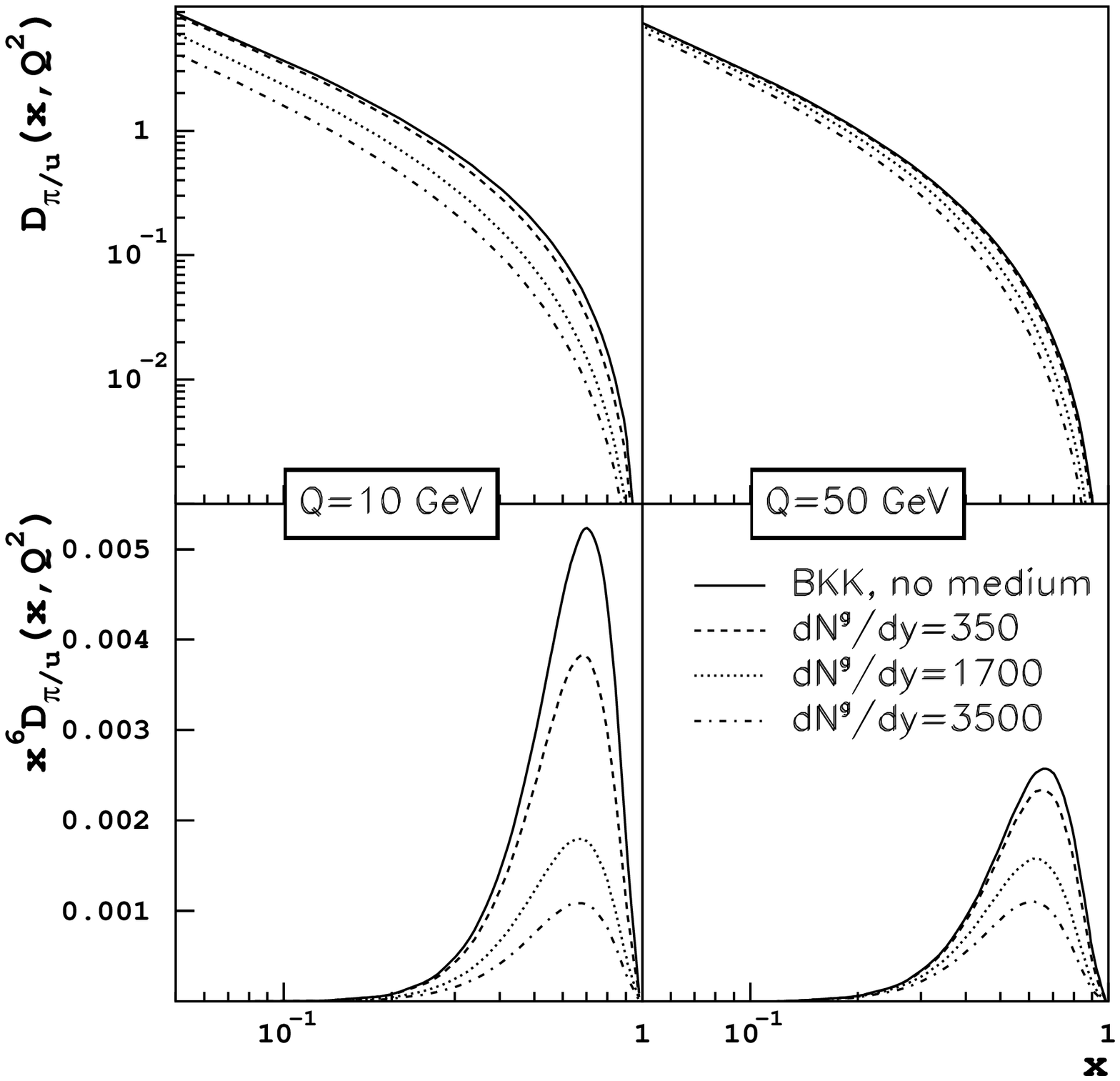, width=70mm}
\caption{Modified fragmentation function for different
gluon densities \cite{salg}}
\label{fig:fig8}
\end{minipage}
\hspace{\fill}
\begin{minipage}[t]{75mm}
\epsfig{bbllx=100,bblly=190,bburx=490,bbury=640,
file=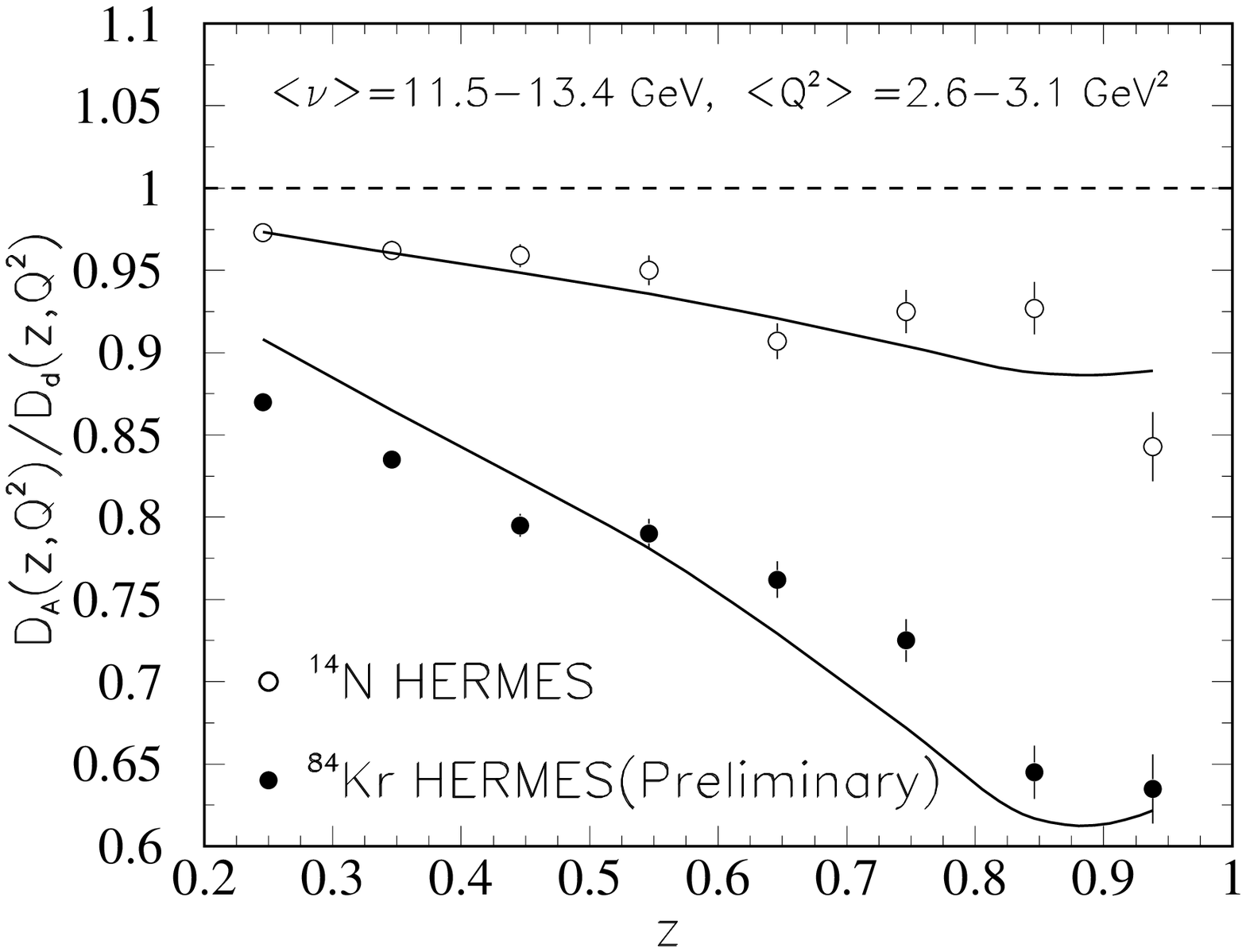, width=60mm}
\caption{Fragmentation function compared with DIS data \cite{wang}}
\label{fig:fig9}
\end{minipage}
\end{figure}

With $D(\E)$ (\ref{prob}) the hadronic cross section
for high $p_{\perp}$ hadrons produced in heavy ion collisions is calculated
by convoluting with the vacuum fragmentation function
\begin{equation}
 D_{h/p}^{\rm medium} (z, Q^2) = \int \, 
{d{\epsilon}} \,
 D(\epsilon = {\hat \epsilon} p_{\perp,jet}) \,
\frac{1}{1-{\hat \epsilon}} \,
 D_{h/p}^{\rm vacuum} (\frac{z}{1-{\hat \epsilon}}\, , Q^2) \, .
\label{dmed}
\end{equation}
As an example the modified fragmentation of $u$ quark $\to \pi$
is plotted in Fig.~\ref{fig:fig8} \cite{salg}.
To $O(\epsilon_{peak}/p_{\perp})$ this ansatz - after convolution with the jet
cross section - is equivalent to (\ref{conv}).

{$\bullet  \, \,$ { effective models}}

Because of the peaked behaviour of $D(\E)$ (Fig.~\ref{fig:fig4})
one may approximate:

 $\, \, \, D(\epsilon) \simeq 
\delta ({\hat \epsilon} - {\hat \epsilon}_{peak})/
p_{\perp,jet}$  
and simplify (\ref{dmed}) by
\begin{equation}
 D_{h/p}^{\rm medium} (z, Q^2) \simeq 
\frac{1}{1- {\hat{\epsilon}_{peak}} }
\, D_{h/p}^{\rm vacuum} (\frac{z}{1- {\hat{\epsilon}_{peak}}} ~, Q^2) \,
\label{approx}
\end{equation}
with ${\hat{\epsilon}_{peak}} ={\epsilon}_{peak}/p_{\perp,jet}$.
Instead, in most cases 
the fractional average loss $\Delta {\hat \epsilon}$, which is larger than
$  {\hat \epsilon}_{peak}$,  is used in (\ref{approx}) for
the phenomenological analysis, e.g. in \cite{vitev,wang}.  

The study of fragmentation functions and their nuclear modifications 
allows a comparison with data in DIS; an example is 
shown in Fig.~\ref{fig:fig9} \cite{wang}.

%
\begin{figure}[htb]
\begin{minipage}[t]{80mm}
\epsfig{file=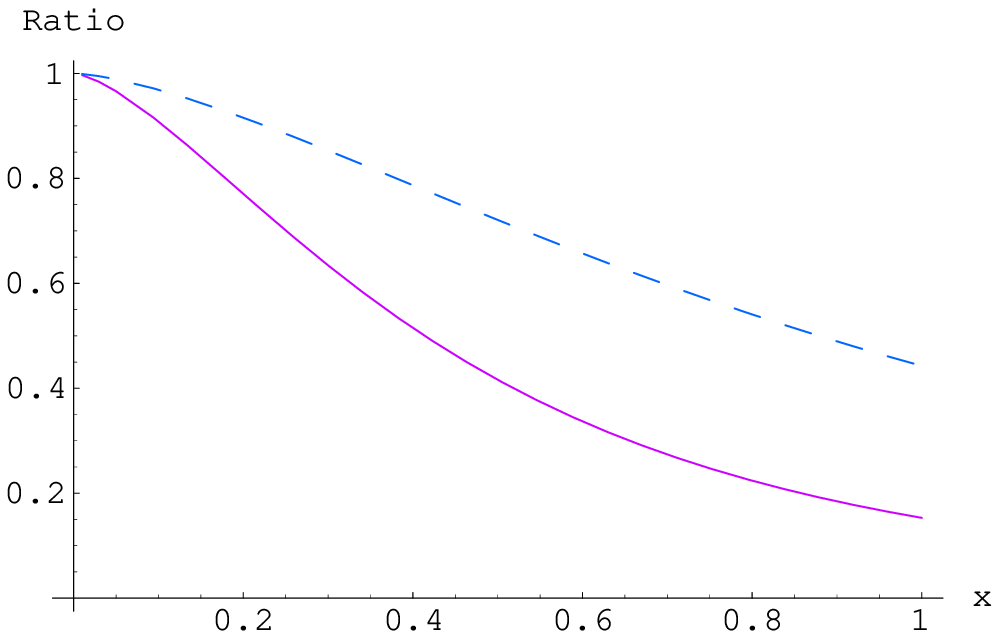,width=70mm}
\caption{Ratio of gluon emission spectra off charm and light quarks
with momenta $p_{\perp} = 10~{\rm GeV}$ (solid) and
$p_{\perp} = 100~{\rm GeV}$ (dashed line)
as a function of $x= \omega/p_{\perp}$ \cite{dima}}
\label{fig:dead}
\end{minipage}
\hspace{\fill}
\begin{minipage}[t]{75mm}
\epsfig{file=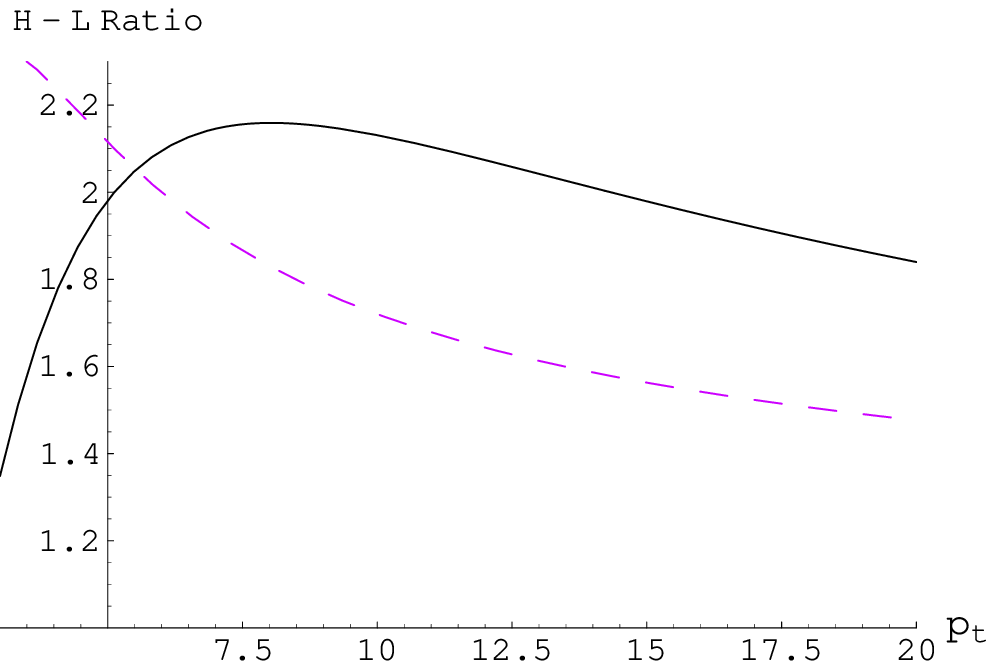,width=70mm}
\caption{Ratio of $Q$ for heavy versus light quarks
in hot matter; dashed curve with $\omega_{cut} = 500~$ MeV \cite{dima}}
\label{fig:fig11}
\end{minipage}
\end{figure}

\subsection{ {Heavy quark radiative energy loss and quenching
\cite{dima} }}

The pattern of medium induced gluon radiation is qualitatively 
different for heavy and light quarks, due to the "dead cone"
phenomena (Fig.~\ref{fig:dead}):
gluon radiation is suppressed at angles $\theta < m_q/E$,
where $m_q$ is the heavy quark mass and $E$ its energy,
\begin{equation}
1/\theta^2  \rightarrow \frac{\theta^2}{[\theta ^2 + m_q^2/E^2 ]^2} \, .
\end{equation}
As a consequence the ratio of quenching factors is
$Q_H(p_{\perp})/Q_L(p_{\perp}) \ge 1$ as shown in Fig.~\ref{fig:fig11},
i.e. there is less energy loss and less suppression for heavy quarks
 than for pions. This effect could be studied experimentally in the $D/\pi$
ratio in heavy ion collisions \cite{dima}. 

\section { {Summary and conclusions} }

After 20 years the studies of hard/large $p_{\perp}$ phenomena
 in nucleus-nucleus collisions 
in the context of { (p)QCD} are now becoming very exciting,
since more and more detailed information is  available.
The importance of multiple gluon scatterings and gluon radiation is
theoretically confirmed. Gluon dynamics in $A-A$ collisions enforces
quenching of large $p_{\perp}$ leading hadrons and jets.
The magnitude is sensitive to the density of the surrounding medium.
Rather stable predictions are possible for $p_{\perp} > 10-20~{\rm GeV}$,
where the quenching factor $Q(p_{\perp})$ is significantly less than $1$.
The most recent observations at RHIC \cite{ref4} are indeed
encouraging (although the measured ratio $R_{AA}(p_{\perp}) \sim Q(p_{\perp})$,
for not yet large values of $p_{\perp} \le 8~{\rm GeV}$,
does not quite show the expected/predicted $p_{\perp}$ dependence
shown in Fig.~\ref{fig:fig5}).

Remaining open questions are related e.g. to the 
determination of the precise properties of the dense medium
(cold hadronic, Colour Glas Condensate, QGP, ..),
and to the relevant time scales: non-equilibrium
 versus thermal and chemical equilibrium.

Detailed quantitative predictions, including 
more complete treatments of  the collision geometry \cite{ref22},
 have still to be worked out.
As reference and  comparison  $p-p$ and $p-A$ data have  to be
taken.

The first indications of the presence of parton energy loss
form a promising start for the further studies of jet physics,
especially for LHC, where  jets with energies of $O(100~{\rm GeV})$
will be measured and where, due to the medium induced gluon radiation,
 a characteristic  dependence
on the finite angular jet cone  is predicted
 \cite{ref17,Yuri,Zakharov,ref20}. 

\vspace{0.5cm}
\noindent
I am grateful for the very pleasant
 collaboration
with Yu.~L.~Dokshitzer, A.~H.~Mueller, S.~Peign\'e and D.~Schiff
during the past few years. I thank U.~A.~Wiedemann for useful
discussions.

\end{document}